\documentstyle[pra,aps]{revtex}

\draft
\input epsf

\begin{document}
\title{Improved characterization of elastic scattering near a Feshbach resonance in $^{85}$Rb}
\author{J.~L. Roberts$^{1}$, James P. Burke, Jr.$^{2}$, N.~R. Claussen$^{1}$, S.~L. Cornish$^{3}$, E.~A. Donley$^{1}$, and C.~E. Wieman$^{1}$}
\address{|$^{1}$JILA,
  National Institute of Standards and Technology and the University of
  Colorado, and
  Department of Physics, University of Colorado, Boulder, Colorado
  80309-0440}
\address{$^{2}$Atomic Physics Division, National Institute of Standards and Technology, Gaithersburg, Maryland 20899-0843}
\address{$^{3}$Current Address:  Clarendon Laboratory, Department of Physics, University of Oxford,
Parks Road, Oxford, OX1 3PU, United Kingdom}
\date{\today}
\maketitle

\begin{abstract}
We report extensions and corrections to the measurement of the Feshbach resonance in $^{85}$Rb cold atom collisions reported earlier [J.~L. Roberts {\it et al.}, Phys. Rev. Lett. {\bf 81}, 5109 (1998)].  In addition to a better determination of the position of the resonance peak (154.9(4) gauss) and its width (11.0(4) gauss), improvements in our techniques now allow the measurement of the absolute size of the elastic scattering rate.  This provides a new measure of the s-wave scattering length as a function of magnetic field near the Feshbach resonance and constrains the Rb-Rb interaction potential.
 \end{abstract}


The presence of a Feshbach resonance in cold atomic collisions allows the elastic and inelastic collision properties to be dramatically altered as an external field is changed \cite{Tiesinga1993}.  In particular, the s-wave scattering length ($a$), which completely characterizes elastic scattering in ultracold collisions, can be tuned over a large range.  In 1998, several groups reported the observation of external-magnetic-field Feshbach resonances in ultracold collisions in $^{85}$Rb \cite{Roberts98,Courteille98} and in $^{23}$Na \cite{Inouye98}.  These tunable interactions have been used to study the stability \cite{Roberts01} and dynamics \cite{Cornish01} of $^{85}$Rb BECs with attractive interactions.  Interpreting the results of these recent studies and other future studies in $^{85}$Rb requires a more accurate knowledge of $a$ as a function of magnetic field ($B$) near the Feshbach resonance than that obtained in Ref. \cite{Roberts98}.  In order to improve the accuracy of the determination of $a$ versus $B$, we extended and improved the measurements performed in Ref. \cite{Roberts98}.

The cold collision Rb-Rb interatomic potential and hence the s-wave scattering length vs. magnetic field is characterized primarily by three parameters:  the triplet scattering length, $a_{T}$; the singlet scattering length, $a_{S}$; and the $C_{6}$ van der Waals coefficient \cite{Burke98,Vogels98}.  In our previous work, we measured two characteristics of the prominent Feshbach resonance in $^{85}$Rb:  the magnetic field at which the magnitude of the s-wave scattering length was maximum ($B_{peak}$) and the magnetic field at which the s-wave scattering length was zero ($B_{zero}$).  This was done by measuring the relative elastic scattering cross section as the magnetic field was changed.  This tightly constrained two of the three parameters, $a_{T}$ and $a_{S}$, for a given $C_{6}$.  However, $C_{6}$ was much less constrained, even when combined with the results from another experiment \cite{Boesten96}.  At the time, the temperature of the trapped atom clouds was too hot to allow the determination of the absolute value of the scattering length \cite{Roberts98}.

We have now been able to improve the measurements made in Ref. \cite{Roberts98} in two ways.  First, the measurement of the stability of BECs with attractive atom-atom interactions (negative $a$) has allowed a much more precise measurement of $B_{zero}$.  This more precise determination showed the presence of some small errors in the previous measurement.  Second, we now have a far greater range of density and temperature of the magnetically trapped atom clouds available, and much better control over those quantities.  The reproducibility and calibration were also improved.  The absolute value of the scattering length could now be accurately determined.  This additional constraint allows us to completely characterize $a$ vs. $B$ near the Feshbach resonance.

The more precise measurement of the position of $B_{zero}$ was a natural outgrowth of our work determining the stability of magnetically trapped BECs with attractive atom-atom interactions \cite{Roberts01}.  When large enough attractive interactions (defined by $Na$) are present, the BECs will become unstable and collapse.  The collapse criterion is given by the relation $N_{cr}|a|=\beta$, where $N_{cr}$ is the number of atoms in the condensate and $\beta=1.41\mu m$ for our magnetic trap.  In Ref. \cite{Roberts01}, $N_{cr}$ was determined as a function of magnetic field down to very small $a$.  A simple extrapolation to the point where $N_{cr}=\infty$ determined the position of $B_{zero}$.  The value of $B_{zero}$ determined in this way is 165.85(5) gauss ($G$), a factor of six more precise than before.  This increase in precision occurs both because the stability measurement is more sensitive and it avoids temperature-dependent shifts.

The new determination of $B_{zero}$ disagrees with the previous measurement \cite{Roberts98} $B_{zero}$=166.8(3)$G$ by more than the stated errors.  A small error in the magnetic field calibration in Ref. \cite{Roberts98} is reponsible for 0.3$G$ of the difference.  The remainder of the difference is likely due to temperature shifts of the elastic cross section and the temperature-dependent asymmetry in the cross section versus magnetic field near $B_{zero}$ \cite{Burkethesis} that were not adequately included in Ref. \cite{Roberts98}.  The new magnetic field calibration also shifts the measured position of $B_{peak}$ to 154.9(4)$G$.

The absolute value of the scattering length was determined at several values of the magnetic field, using ``cross-dimensional mixing'' measurements as in Ref. \cite{Roberts98}.  In this technique, a cloud of cold atoms, created by evaporative cooling as in Ref. \cite{Cornish00}, is perturbed from its equilibrium energy distribution.  The time for the cloud to equilibrate by elastic collisions is then measured.  In the zero-temperature limit, the elastic collision cross section ($\sigma$) is $\sigma=8\pi a^{2}$.

Once the atom cloud had been cooled sufficiently, the bias magnetic field in the trap was ramped to near $B_{zero}$ to drastically reduce the elastic collision rate.  By cutting faster with the rf-``knife'' used for evaporative cooling than the cloud could equilibrate, it was possible to remove energy in just the radial direction of the axially symmetric magnetic trap \cite{Roberts98}.  The width of the cloud in the radial direction was decreased until the radial energy had been reduced to 0.6 of the axial direction, and then the bias magnetic field was ramped adiabatically to a selected value and the cloud was allowed to equilibrate for a fixed time.  The size, number, and shape of the cloud were measured using on-resonance destructive absorption imaging.  This procedure was repeated for different equilibration times and the rate of the relaxation of the aspect ratio of the clouds (the ratio of the axial width to the radial width) to equilibrium was recorded.

The rate of flow of energy out of any particular direction, axial or radial, is proportional to the difference between the average energy in that direction and the equilibrium value multiplied by the factor $\Gamma=\frac{2}{\kappa}\langle n\rangle \sigma \langle v\rangle$ (i.e. $\frac{dE_{i}}{dt}=-\Gamma (E_{i}-E_{avg})$, i=x,y,z).  In the factor $\Gamma$, $\langle n\rangle$ is the average density defined by $\frac{\int n(\vec{x})^{2}d^{3}x}{\int n(\vec{x})d^{3}x}$, $\langle v\rangle$ is the root-mean-square relative speed $\sqrt{\frac{16k_{B}T}{\pi m}}$, $\sigma=8\pi a^{2}$ is the elastic collision cross section, and $\kappa$ is a numerical factor.
The parameter $\kappa$ is weakly dependent on the difference between radial and axial energy but is always within 2\% of the value 2.50.  It is calculated using classical transport theory \cite{Reif} with the assumption that the velocity and density distributions can be described in each direction by a single parameter, an effective temperature that is equal to the average energy divided by the Boltzmann constant $k_{B}$.  A Monte-Carlo simulation agrees with this classical calculation at the 4\% level \cite{Demarco99}.

The relaxation of the aspect ratio was fit appropriately to determine $\Gamma$ and hence $a$.  $\Gamma$ is not strictly constant as a function of equilibration time since the density and temperature of an atom cloud varied during equilibration due to loss, heating, and change in shape.  These small ($<5\%$) variations were included in the fits to the data.  The density and temperature of the atom cloud was measured directly along with the aspect ratio in the absorption imaging.  The density was maintained near $2.5\times 10^{10} $cm$^{-3}$ and the temperature was kept near 130 nK.  For these cold clouds, the energy dependence of the average elastic cross section is expected to be less than 5\%, and the fits to the data included the energy dependence due to the unitarity limit \cite{unitarity}.

Data were taken at $B$=168.0, 169.7, and 251.0 $G$.  The magnetic field was calibrated in the same way as Ref. \cite{Roberts98}:  the rf frequency that resonantly drove spin flip transitions for atoms at the center of the cloud was measured and the Breit-Rabi equation was used to determine the magnetic field.  Colder clouds were used, improving the precision.  The spread of the magnetic field across the clouds was $\sim 0.2 G$ FWHM.  Figure 1 shows a set of aspect ratio equilibration data and the corresponding fit.  Across the range of magnetic fields measured, the equilibration times varied by about a factor of 30.  The distance from $B_{peak}$ ensures that the spread in $a$ due to the spread in $B$ across the cloud was not significant and that the precise location of $B_{peak}$ is also not a limiting factor in analyzing the results.

The uncertainty in the determination of $a$ is dominated by the determination of number and the fit to obtain $\Gamma$.  The number determination relies on the calibration of our absorption imaging system, which was performed while taking into account the frequency and intensity of our probe laser, optical pumping effects, and the small Doppler shift arising from scattering multiple photons.  We estimate the uncertainty to be 10\%, which results in a 5\% uncertainty in each determination of $a$.  The uncertainty in the fit to obtain $\Gamma$ is primarily caused by small oscillations in the aspect ratio due to the vertical and horizontal radial directions not being perfectly degenerate.  Although this limited the accuracy of the fit, the size of this imbalance was varied and no significant shifts in the results were observed.  Also, the period of the oscillations does not depend on the value of $a$ and so the agreement between data sets of very different equilibration times also indicates that this effect is not distorting the results significantly.  The measured value (in atomic units) of $a$ is -342(10), -97(6), and -60(4) at $B$ = 251.0, 169.7, and 168.0$G$, respectively.  The listed uncertainty does not include the common uncertainty due to the atom number determination.

Near the Feshbach resonance, the change in $a$ with $B$ can be well approximated by the relation
$a=a_{bg}(1-\frac{B_{zero}-B_{peak}}{B-B_{peak}})$ \cite{approximation}.  Using this, we can determine $a_{bg}$ for each measured value of $a$ (Fig. 2).  The weighted average of all of these data gives $a_{bg}=-380(21)$ where the uncertainty is dominated by the determination of number.

Our measurement now constrains all three parameters needed to describe the Rb-Rb potential for ultracold collisions.  Using the same methods presented in Ref. \cite{Roberts98}, the Rb-Rb Born-Oppenheimer potentials are adjusted until the theoretically predicted $B_{zero}$, $B_{peak}$, and $a_{bg}$ match the observed values.  The calculation of $a$ vs. $B$ is performed using a quantum defect method \cite{Burke98a}.  Matching the Rb-Rb potential to the observed scattering lengths implies $a_{T}$($^{85}$Rb) = $-332\pm 18$, $a_{S}$($^{85}$Rb)=$3650^{+1500}_{-670}$ \cite{singlet}, and $C_{6}=4660\pm 20$, where all of the values are listed in atomic units \cite{Marinescu98}.  This value of the $C_{6}$ coefficient is consistent both with an {\it ab initio} calculation \cite{Deverianko99} and a recent photoassociation measurement \cite{Vogels00}.

As a result of improvements in our experimental techniques and the measurement of $^{85}$Rb BEC stability in the presence of attractive interactions, we have been able to improve our characterization of the prominent Feshbach resonance in $^{85}$Rb.

This work has been supported by the ONR and NSF.  We are pleased to acknowledge helpful discussions with Brian DeMarco and Murray Holland.  One of us (S.~L. Cornish) acknowledges the support of the Lindemann Foundation.

\begin{figure}
\caption{Relaxation of the aspect ratio to equilibrium.  The data shown here were taken at $B$=169.7$G$.  Each data point ($\bullet$) represents a single destructive absorption measurement.  The solid line shows the fit to the data used to determine the elastic scattering cross section and hence the scattering length.  The deviation of the data points from the fit around 3-4 seconds is due to the oscillations between the two radial directions as described in the main text.}
\end{figure}

\begin{figure}
\caption{The value of $a_{bg}$ implied by several cross-dimensional mixing measurements.  This value was determined by dividing the s-wave scattering length determined at a particular magnetic field by the factor $(1-\frac{B_{zero}-B_{peak}}{B-B_{peak}})$.  The s-wave scattering lengths measured at 251.0$G$ were divided by an additional factor of 0.99 to account for the deviation of the approximate description of $a$ vs. $B$ from the full theory calculation.  The error bars shown include the statistical errors plus the fit uncertainty due the initial radial energy imbalance.  The fit uncertainty is non-statistical and so the central values are clustered together more than the error bars would suggest.}
\end{figure}


\noindent

\begin{references}

\bibitem{Tiesinga1993} W.~C. Stwalley, Phys. Rev. Lett. {\bf 37}, 1628 (1976); E. Tiesinga, A. Moerdijk, B.~J. Verhaar, and H.~T.~C. Stoof, Phys. Rev. A {\bf46}, R1167 (1992).

\bibitem{Roberts98}  J.~L.~Roberts {\it et al.}, Phys. Rev. Lett., {\bf 81}, 5109 (1998).

\bibitem{Courteille98} Ph. Courteille {\it et al.}, Phys. Rev. Lett. {\bf 81}, 69 (1998).

\bibitem{Inouye98} S. Inouye {\it et al.}, Nature (London) {\bf 392}, 151 (1998).

\bibitem{Roberts01} J.~L. Roberts {\it et al.}, submitted to Phys. Rev. Lett.

\bibitem{Cornish01} S.~L. Cornish {\it et al.}, submitted to Science.

\bibitem{Burke98} J.~P. Burke, Jr., C.~H. Greene, and J.~L. Bohn, Phys. Rev. Lett. {\bf 81}, 3355 (1998).

\bibitem{Vogels98} J.~M. Vogels, B.~H. Verhaar, and R.~H. Blok, Phys. Rev. A {\bf 57}, 4049 (1998).

\bibitem{Boesten96} H.~M.~J.~M. Boesten {\it et al.} Phys. Rev. Lett. {\bf 77}, 5194 (1996).

\bibitem{Burkethesis} J.~P. Burke, Jr., Ph. D. thesis, University of Colorado, 1999.

\bibitem{Cornish00}  S.~L. Cornish {\it et al.}, Phys. Rev. Lett. {\bf 85}, 1795 (2000).

\bibitem{Reif} See, for instance, F. Reif, {\it Fundamentals of Statistical and Thermal Physics}, (McGraw Hill, New York, 1965), pp. 516-527.

\bibitem{Demarco99} B. DeMarco {\it et al.}, Phys. Rev. Lett. {\bf 82}, 4208 (1999).

\bibitem{unitarity}  For instance, see J. R. Taylor, {\it Scattering Theory: The Quantum Theory of Nonrelativistic Collisions} (Kreiger, Malabar, Florida, 1987), p. 89.

\bibitem{approximation} From 155.3$G$ ($a \sim 550$ nm) to 175$G$, this agrees with the full theory (see [10]) to better than 0.5\%, and at 251$G$ there is a 1\% difference.

\bibitem{Burke98a}  J.~P. Burke, Jr., C.~H. Greene, and J.~L. Bohn, Phys. Rev. Lett. {\bf 81}, 3355 (1998).

\bibitem{singlet} The singlet scattering potential has a bound state very near to the collision threshold and so $a_{S}$ is close to diverging.  This accounts for the large error bars on $a_{S}$.

\bibitem{Marinescu98}  The quoted values for $a_{T}, a_{S}$ and $C_{6}$ are calculated using the values of $C_{8}$ and $C_{10}$ predicted in M. Marinescu and L. You, Phys. Rev. Lett. {\bf 81} 4596 (1998).  Allowing C$_{8}$ to vary by 10\% would increase the uncertainties on $a_{triplet}, a_{singlet}$ and C$_{6}$ by about 25\%.

\bibitem{Deverianko99}  A. Derevianko {\it et al.}, Phys. Rev. Lett. {\bf 82}, 3589 (1999).  In this work, $C_{6}$ was calculated to be 4691$\pm 50$.

\bibitem{Vogels00}  J.~M. Vogels {\it et al.}, Phys. Rev. A {\bf 61}, 043407 (2000).  Here, $C_{6}$ was measured to be 4650$\pm 50$.

\end{references}
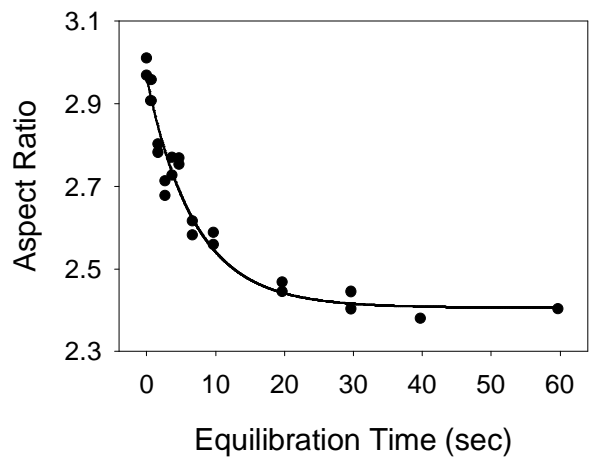
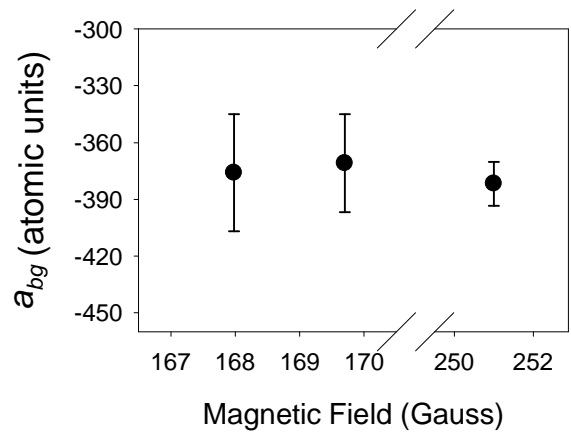
\end{document}